\documentclass[11pt]{article}
\usepackage{comment}
\usepackage[hidelinks]{hyperref}
\usepackage{enumitem}
\usepackage{amsmath,amssymb,epsf,cite,graphicx,subfigure,physics}
\usepackage{amsmath,amsthm, tikzsymbols, dsfont, amsfonts}
\usepackage{algorithm}
\usepackage{algorithmic}
\usepackage{forloop}
\usepackage{mathtools}
\usepackage{comment}
\usepackage{xcolor}
\usepackage{graphicx}
\usepackage{dsfont}
\usepackage{soul}

\allowdisplaybreaks

\numberwithin{equation}{section}

\newcommand{\zone}{z_1, \bar{z}_1}
\newcommand{\ztwo}{z_2, \bar{z}_2}

\newcommand{\qone}{q(\zone)}
\newcommand{\qtwo}{q(\ztwo)}
\newcommand{\vqone}{\vec{q}(\zone)}
\newcommand{\vqtwo}{\vec{q}(\ztwo)}

\definecolor{airforceblue}{rgb}{0.36, 0.54, 0.66}
\newcommand{\beq}{\begin{equation}}
\newcommand{\eeq}{\end{equation}}

  \theoremstyle{definition}

\newcommand{\bz}{{\bar{z}}}

\newcommand{\J}{\mathsf{J}}
\newcommand{\K}{\mathsf{K}}
\newcommand{\sP}{\mathsf{P}}
\newcommand{\sR}{\mathsf{R}}
\newcommand{\sL}{\mathsf{L}}
\newcommand{\bL}{\bar{\mathsf{L}}}

\newcommand{\sS}{\mathsf{S}}

\newcommand{\rsw}{\mathsf{RSW}}

\newcommand{\drsw}{{\dagger_{\mathsf{RSW} }}}

\newcommand{\nn}{n}
\newcommand{\nm}{m}
\newcommand{\np}{p}
\newcommand{\nl}{k}

\newcommand{\mj}{\mathfrak{j}}
\newcommand{\tj}{\widetilde{\mathfrak{j}}}

\begin{document}
\baselineskip=15.5pt
\pagestyle{plain}
\setcounter{page}{1}
\begin{center}
{\LARGE \bf An Integer Basis for Celestial Amplitudes}
\vskip 1cm

\textbf{Jordan Cotler$^{1,2,a}$, Noah Miller$^{2,b}$, Andrew Strominger$^{1,2,c}$}

\vspace{0.5cm}

$^{1}${\it Harvard Society of Fellows, Cambridge, MA 02138 USA \\}
$^{2}${\it Center for Fundamental Laws of Nature, Harvard University, Cambridge, MA 02138 USA \\}

\vspace{0.3cm}

{\tt ${}^a$jcotler@fas.harvard.edu, ${}^b$noahmiller@g.harvard.edu, ${}^c$andrew$\_$strominger@harvard.edu}

\medskip

\end{center}

\vskip1cm

\begin{center}
{\bf Abstract}
\end{center}
\hspace{.3cm} 
We present a discrete basis of solutions of the massless Klein-Gordon equation in $3+1$ Minkowski space which transform as $\mathfrak{sl}(2,\mathbb{C})$ Lorentz/conformal primaries and  descendants, and whose elements all have integer conformal dimension.  We show that the basis is complete in the sense that the Wightman function can be expressed as a quadratic sum over the basis elements.

\newpage

\tableofcontents

\section{Introduction}

Celestial amplitudes are ordinary quantum field theory or quantum gravity scattering amplitudes re-expressed in a basis of $\mathfrak{sl}(2, \mathbb{C})$ Lorentz/conformal primary scattering states, which can be viewed as operators on the celestial sphere.  The set of all conformal primary wavefunctions is vastly overcomplete. Finding an optimal complete basis is a central problem in celestial holography. It is similar to the now-solved problem~\cite{Zamolodchikov:1995aa, Teschner:2001rv} of finding  a complete basis of operators in 2D Liouville theory.  Because CCFTs are not unitary, it is not obvious what properties an optimal basis should have. Some natural ones are:
    \begin{itemize}
    \item Forming irreducible representations of the Poincar\'{e} group;
    \item Instantiating closure, locality and associativity of the celestial OPE;
    \item Providing a factorization of scattering amplitudes onto associated 2D conformal blocks; and
     \item Providing a solution of the celestial conformal bootstrap equation.
     \end {itemize}
 One frequently considered basis is the unitary principal series \cite{pasterski2017conformal}. This is complete on the space of square normalizable wavefunctions, but does not appear to satisfy all of the above constraints.  Other possibilities involve restricting the conformal weights to integer values. This was put forth in \cite{Atanasov:2021oyu} based on periodicity conditions on the celestial torus.  Other works have also studied integer conformal dimensions in the context of large gauge symmetry in the spin-1 and spin-2 cases~\cite{Kapec:2016jld, Donnay:2018neh,  Puhm:2019zbl, Fan:2019emx, Guevara:2021abz, Adamo:2019ipt, Donnay:2020guq}, twisted holography~\cite{Costello:2022wso}, as well as more broadly~\cite{Karateev:2017jgd, Donnay:2022wvx,
 Brown:2022miw, Sun:2021thf, Atanasov:2021cje, Guevara:2022qnm, Banerjee:2019aoy, Banerjee:2019tam, Kapec:2021eug, Jorge-Diaz:2022dmy, Banerjee:2022wht, Ball:2022bgg, Pasterski:2021fjn}.
 
In this paper we construct a complete integer basis employing irreducible integer representations of the Poincar\'{e} group. 
 As our work was nearing completion, a complete integer basis for the Schwartz space of solutions of the massless scalar wave equation  was found in \cite{Freidel:2022skz}. These results contain overlap with ours, although the details and methods are quite different.
 
 Our approach was inspired by earlier work on the completeness of quasinormal mode bases~\cite{Berti:2009kk, ching1996wave, beyer1999completeness, Nollert:1998ys, Jafferis:2013qia}.  We draw particular attention to~\cite{Jafferis:2013qia}, in which it is shown that there exists an integer basis of quasinormal modes for de Sitter space which are complete in the sense that the Wightman function can be represented as a sum over such modes. In the flat limit of de Sitter, the quasinormal modes decompose as tower of integer conformal primaries transforming irreducibly under Poincar\'{e}.  This suggests a similar integer basis of conformal primaries should exist for Minkowski space, as is demonstrated herein.\footnote{Although we do not herein attempt to make the limiting connection with de Sitter precise.} 

 Following~\cite{Jafferis:2013qia}, our strategy is to find a mode decomposition of the Wightman function $G_+(X,Y)$ in terms of an integer basis of solutions to the massless Klein-Gordon equation (i.e.~the wave equation).  Reproducing the Wightman function is sufficient to establish that our basis can propagate arbitrary initial data to solutions of the the massless Klein-Gordon equation, as we explain in the Appendices.  Along the way, we discover a novel modification of the Klein-Gordon inner product which emulates that of BPZ~\cite{belavin1984infinite}, and has had recent manifestations in celestial holography~\cite{Crawley:2021ivb}.

 The remainder of the paper is organized as follows.  In Section~\ref{sec_integer_bases} we define our integer basis.  In Sections~\ref{sec_rsc} and~\ref{sec_adjoints} we discuss a modification to the Klein-Gordon inner product which manifests our basis.  In Section~\ref{sec_decomp} we give our main result, namely a decomposition of the Wightman function into our integer basis.  We conclude with a brief discussion in Section~\ref{sec_discussion}.  Following this are numerous Appendices; of particular note is Appendix~\ref{App:finitedim} which explains how to organize our integer basis into finite-dimensional representations of the Lorentz group.

\section{Candidate integer basis}\label{sec_integer_bases}

In this section we present an `integer basis' of solutions to the massless Klein-Gordon equation, in which each element has integer conformal weight. The basis is composed of a shadow pair of two irreducible representations of the Poincar\'{e} group, both  of which decompose into towers of finite-dimensional representations of the $\mathfrak{sl}(2, \mathbb{C})$ Lorentz/conformal group. In subsequent sections we will show that our integer bases are complete in the sense that they provide a mode expansion for the Wightman function. We use the ($-$$+$$+$$+$) metric convention in this work.

Defining the Poincar\'{e} generators as
\begin{align}\label{poincare_generators}
    \J_i = -\epsilon_{ijk} X^j \partial_k\,, \hspace{1 cm} \K_i = -X^0 \,\partial_i - X^i\,\partial_0\,, \hspace{1 cm} \sP_\mu = -\partial_\mu\,,
\end{align}
where $i = 1,2,3$, we have the non-zero commutators
\begin{align}
    [\J_i, \J_j] &= \epsilon_{ijk}\,\J_k\,, \qquad
    [\J_i, \K_j] = \epsilon_{ijk}\,\K_k\,,\qquad
    [\K_i, \K_j] = -\epsilon_{ijk}\,\J_k\,, \\
    [\K_i, \sP_0] &= \sP_j\,, \qquad \quad \,\,\, [\K_i, \sP_i] = \sP_0\,, \qquad \quad \,\,\,\,\,\, [\J_i, \sP_j] = \epsilon_{ijk}\,\sP_k\,.\nonumber
\end{align}
We recall that the Lorentz algebra $\mathfrak{so}(1,3)$ is isomorphic to the 2D global conformal transformations $\mathfrak{sl}(2, \mathbb{C})$.  The identification can be made explicit via
\begin{equation}\label{sl2c_generators}
  \begin{split}
    \sL_0  &= \frac{1}{2} (-\K_3 - i \J_3)\,, \\
    \sL_{1} &= \frac{1}{2} (-\K_1 + \J_2 - i (\K_2 + \J_1) )\,,  \\
    \sL_{-1} &= \frac{1}{2} (\K_1 + \J_2 - i (\K_2 - \J_1) )\,,
  \end{split}
\quad \quad \quad
  \begin{split}
    \bL_0  &= \frac{1}{2} (-\K_3 + i \J_3)\,,  \\
    \bL_{1} &= \frac{1}{2} (-\K_1 + \J_2 + i (\K_2 + \J_1 ) )\,,  \\
    \bL_{-1}  &= \frac{1}{2} (\K_1 + \J_2 + i (\K_2 - \J_1) )\,,
  \end{split}
\end{equation}
which satisfy the usual commutation relations
\begin{equation}\label{L_commutators}
    [\sL_m, \sL_n] = (m-n) \sL_{m+n}\,, \hspace{1 cm} [\bL_m, \bL_n] = (m-n) \bL_{m+n}\,,  \hspace{1 cm} 
    [\sL_m, \bL_n] = 0\,.
\end{equation}

To construct a basis of solutions to the massless Klein-Gordon equation, we start by identifying solutions which transform as 2D conformal primaries.  There are two kinds of such solutions, given by~\cite{Pasterski:2016qvg,pasterski2017conformal}
\begin{align}
\label{E:twokinds1}
    \psi_{\Delta, q}(X) &= \frac{1}{(- q \cdot X)^\Delta}\,,\qquad \widetilde{\psi_{\Delta, q}}(X) = \frac{(-X^\mu X_\mu)^{1 - \Delta}}{(- q \cdot X)^{2-\Delta}}\,,
\end{align}
where $q^\mu$ is a null vector satisfying $q^2 = 0$.  For our purposes, we will always take $\Delta$ to be an integer. To verify that these functions transform as primaries, we parameterize the null vector with a complex number $z$ as
\begin{equation}
    q^\mu(z, \bz) = (1 + z \bz, z + \bz, -i(z - \bz), 1 - z \bz)
\end{equation}
and directly compute
\begin{align}\label{primary_Ln_1}
    \sL_n \psi_{\Delta, q(z, \bz)} &= \left( \frac{\Delta}{2} (n+1) z^n + z^{n+1} \partial_z \right) \psi_{\Delta, q(z, \bz)}  \\
    \label{primary_Ln_2}
    \bL_n \psi_{\Delta, q(z, \bz)} &= \left( \frac{\Delta}{2} (n+1) \bz^n + \bz^{n+1} \partial_\bz\right) \psi_{\Delta, q(z, \bz)} \\
    \label{shadow_Ln_1}
    \sL_n \widetilde{\psi_{\Delta, q(z, \bz)} }&= \left( \frac{(2 - \Delta)}{2} (n+1) z^n + z^{n+1} \partial_z\right) \widetilde{\psi_{\Delta, q(z, \bz)}} \\
 \label{shadow_Ln_2}
    \bL_n \widetilde{\psi_{\Delta, q(z, \bz)} } &= \left( \frac{(2 - \Delta)}{2} (n+1) \bz^n + \bz^{n+1} \partial_\bz\right) \widetilde{\psi_{\Delta, q(z, \bz)}}
\end{align}
for $n = -1,0,1$. From the above, we see that each $\psi_{\Delta, q(z, \bz)}$ transforms as conformal primary of weight $\Delta$ and $(h, \bar{h}) = (\frac{\Delta}{2}, \frac{\Delta}{2})$, and similarly each $\widetilde{\psi_{\Delta, q(z, \bz)} }$ transforms as a conformal primary of weight $2-\Delta$ and $(h, \bar{h}) = (\frac{2-\Delta}{2}, \frac{2-\Delta}{2})$.

To formulate a candidate basis of solutions to the massless Klein-Gordon equation, we would like to find solutions which are highest weight with respect to the $\mathfrak{sl}(2,\mathbb{C})$ algebra so that we can fill out our basis by taking descendants.  For this purpose, let us note a convenient choice of $q^\mu$, namely a null vector which points towards the north pole:
\begin{align}
    N^\mu \vcentcolon= q^\mu(0,0) = (1,0,0,1)\,.
\end{align}
From~\eqref{primary_Ln_1},~\eqref{primary_Ln_2},~\eqref{shadow_Ln_1},~\eqref{shadow_Ln_2}, we have 
\begin{align}
    \sL_{1} \psi_{\Delta, N} &= \bL_{1} \psi_{\Delta,N} = 0 \\
    \sL_0 \psi_{\Delta, N} &= \bL_0 \psi_{\Delta, N} = \frac{\Delta}{2} \,\psi_{\Delta, N} \\
    \sL_{1} \widetilde{\psi_{\Delta, N}} &= \bL_{1} \widetilde{\psi_{\Delta,N}} = 0 \\
    \sL_0 \widetilde{\psi_{\Delta, N}} &= \bL_0 \widetilde{\psi_{\Delta, N}} = \frac{2-\Delta}{2} \,\widetilde{\psi_{\Delta, N}}\,,
\end{align}
thus establishing that $\psi_{\Delta, N}$ and $\widetilde{\psi_{\Delta, N}}$ are highest weight states for any (integer) $\Delta$.  We can then form $\mathfrak{sl}(2, \mathbb{C})$ descendants given by $\sL^n_{-1} \bL^{\bar{n}}_{-1} \psi_{\Delta, N}$ and $\sL^n_{-1} \bL^{\bar{n}}_{-1} \widetilde{\psi_{\Delta, N}}$, for $n$ and $\bar{n}$ distinct non-negative integers.  A similar analysis shows that $\psi_{\Delta, S}$ and $\widetilde{\psi_{\Delta, S}}$ are lowest-weight states, where 
\begin{equation}
    S^\mu \vcentcolon= (1,0,0,-1)
\end{equation}
is a null vector which points towards the south pole.

With the above ingredients at hand, we are almost prepared to propose a candidate basis.  Let us define the positive ($+$) and negative ($-$) frequency wavefunctions~\cite{pasterski2017conformal}
\begin{align}
 \psi^\pm_{\Delta, N}(X) \vcentcolon= \psi_{\Delta, N}(X_\pm)\,,\qquad \widetilde{\psi^\pm_{\Delta, N}}(X) \vcentcolon= \widetilde{\psi_{\Delta, N}}(X_\pm)\,,
\end{align}
where
\begin{equation}
    X^\mu_\pm \vcentcolon= (X^0 \mp i \epsilon, X^1, X^2, X^3)\,.
\end{equation}
We further define the reflection operator $\textsf{R}$ which takes 
\begin{equation}
\mathsf{R} : (X^0, X^1, X^2, X^3) \longmapsto (X^0,X^1,X^2,-X^3)\,,
\end{equation}
and use $\textsf{R} f(X)$ to denote $f(\textsf{R}X)$. The operator is useful since it interchanges highest-weight and lowest-weight states:
\begin{equation}
\textsf{R} \psi_{\Delta,N}^\pm = \psi_{\Delta,S}^\pm\,,\qquad \textsf{R} \widetilde{\psi_{\Delta,N}^\pm} = \widetilde{\psi_{\Delta,S}^\pm}\,.
\end{equation}

Defining the sets
\begin{align}
\label{E:Cset1}
\mathcal{C} &\vcentcolon=  \Big\{\sL_{-1}^n \bL_{-1}^{\bar{n}} \psi_{\Delta, N}(X_+)\Big\}_{\substack{n,\bar{n}=0,1,..., -\Delta \\ \!\!\Delta = 0,-1,-2,...}} \\
\label{E:Cset2}
\widehat{\mathcal{C}} &\vcentcolon= \Big\{\sL_{1}^n \, \bL_{1}^{\bar{n}} \,\widetilde{\psi_{2-\Delta, S}}(X_-)\Big\}_{\substack{n,\bar{n}=0,1,..., -\Delta \\ \!\!\Delta = 0,-1,-2,...}}\,,
\end{align}
we are now fully equipped to propose the integer basis
\begin{align} \label{basis_1}
\mathcal{B} \vcentcolon= \mathcal{C} \cup \widehat{\mathcal{C}}\,.
\end{align}
Notice that if $\mathcal{B}$ is an integer basis, then the complex conjugate of $\mathcal{B}$, namely $\mathcal{B}^* = \mathcal{C}^* \cup \widehat{\mathcal{C}}^*$, is likewise an integer basis.

\section{The $\rsw$ inner product}\label{sec_rsc}

There is a useful relationship between the sets $\mathcal{C}$ and $\widehat{\mathcal{C}}$ defined above: they are related by an ``$\rsw$'' transformation, where $\sR$, $\sS$, and $\mathsf{W}$ are independent linear operators. Having already defined $\sR$, we define the shadow operator $\textsf{S}$ which acts on the function $\psi_{\Delta,q(z,\bar{z})}^\pm$ with conformal dimension $\Delta$ by\footnote{Setting $z = x + i y$ we use the conventions $d^2 z \vcentcolon= dx \, dy$ and $\delta^2(z) \vcentcolon= \delta(x) \delta(y)$. } 
\begin{equation}\label{S_def}
    \sS \psi^\pm_{\Delta, q(z, \bar{z})} \vcentcolon= \frac{\Delta - 1}{\pi} \int d^2 w \,\frac{1}{|z-w|^{2(2 - \Delta)}} \,\psi^\pm_{\Delta, q(w, \bar{w})}\,.
\end{equation}
The shadow operator relates our two kinds of primary states in~\eqref{E:twokinds1} by \cite{simmons2014projectors, pasterski2017conformal}
\begin{equation}
    \sS \psi^\pm_{\Delta, q(z, \bar{z})} = \widetilde{\psi_{\Delta, q(z, \bz)}^\pm}\,.
\end{equation}
We also define the $\mathsf{W}$ operator by
\begin{equation}
    \mathsf{W} \psi^\pm_{\Delta,q} \vcentcolon= \psi^\pm_{2 - \Delta,q}\,,
\end{equation}
which simply replaces the conformal weight $\Delta$ by $2 - \Delta$. 

Having defined $\mathsf{R}$, $\mathsf{S}$, and $\mathsf{W}$ individually, we now define the $\rsw$ inner product.  Recall that the Klein-Gordon inner product is
\begin{equation}
\label{E:KGnorm1}
(f,g)_{\mathrm{KG}} \vcentcolon= i \int_{\Sigma} d^3 X \, (f^* \partial_0 g - g \,\partial_0 f^*)\,,
\end{equation}
where $\Sigma$ is a Cauchy slice at fixed $X^0$.  If $f(X)$ and $g(X)$ satisfy the free Klein-Gordon equation (i.e.~the ordinary wave equation) $\square f = \square g = 0$, then~\eqref{E:KGnorm1} is independent of the Cauchy slice $\Sigma$ on which the integral is evaluated.  Then the $\rsw$ inner product is defined as
\begin{equation}
\begin{split}
    \big( f, g \big)_{\rsw} \vcentcolon= \big( f, (\rsw) g \big)_{\rm KG}\,.
\end{split}
\end{equation}
This is closely related to the shadow product at opposite poles introduced in \cite{Crawley:2021ivb}. It should be noted that $\rsw$ inner products of our integer bases are finite numbers. For instance, in Appendix \ref{app_inner_prod} we compute the inner product of two primary states $\psi^\pm_{\Delta, q(z, \bz)}$ and find it to be
\begin{equation}\label{rsw_primary}
    \left( \psi^\pm_{\Delta_1, q(z_1, \bz_1)}, \psi^\pm_{\Delta_2, q(z_2, \bz_2)} \right)_{\rsw} = \mp ( 2 \pi)^2 |1 - z_1 \bz_2 |^{-2\Delta_1} \delta_{\Delta_1, \Delta_2}.
\end{equation}

We can further define the `$\rsw$-adjoint' of wave functions, denoted $\drsw$, via
\begin{equation}
\label{E:fdag}
f^{\dagger_\rsw} \vcentcolon= (\rsw) f^* \, , \hspace{1 cm} \big(f, g\big)_\rsw = \big((g^*)^{\dagger_\rsw}, (f^*)^{\dagger_\rsw}\big)_\rsw^*\,.
\end{equation}
It can be readily shown that $\sR^2 = 1$, $(\mathsf{SW})^2 = 1$, and $[\sR, \mathsf{SW}] = 0$. These imply that
\begin{equation}
    (\rsw)^2 = 1
\end{equation}
so $(f^\drsw)^\drsw = f$. In fact, the $\rsw$ adjoint takes wave functions from $\mathcal{C}$ into $\widehat{\mathcal{C}}$, and vice versa:
\begin{equation}\label{rsw_descendants}
    \Big( \sL_{-1}^n \bL_{-1}^{\bar{n}} \psi_{\Delta, N}(X_+) \Big)^\drsw = \sL_{1}^n \bL_{1}^{\bar{n}} \widetilde{\psi_{2-\Delta, S}}(X_-).
\end{equation}
This relation will be proven in the following section.

\section{$\rsw$ adjoints}
\label{sec_adjoints}

The Klein-Gordon adjoint $\dagger$ of a linear operator $\mathcal{O}$ is defined by $(f, \mathcal{O}^\dagger g)_{\rm KG} = ( \mathcal{O} 
 f,g)_{\rm KG}$. Likewise, we can define the $\rsw$ adjoint $\drsw$ by
\begin{equation}
    \mathcal{O}^\drsw \vcentcolon= ( \rsw) \mathcal{O}^\dagger (\rsw) \; , \hspace {1 cm} \big(f,  \mathcal{O}^{\dagger_\rsw}  g\big)_\rsw = \big(\mathcal{O} f,  g\big)_\rsw.
\end{equation}
Let us investigate the $\rsw$ adjoint properties of the $\mathfrak{sl}(2, \mathbb{C})$ generators. Note that our Poincar\'{e} generators \eqref{poincare_generators} are skew adjoint under the KG adjoint, satisfying
\begin{equation}
    \J_i^\dagger = - \J_i, \hspace{1 cm} \K_i^\dagger = - \K_i, \hspace{1 cm} \sP_\mu^\dagger = - \sP_\mu.
\end{equation}
From this, the $\mathfrak{sl}(2, \mathbb{C})$ generators \eqref{sl2c_generators} then inherit the Klein-Gordon adjoint property
\begin{equation}\label{Ln_kg_adj}
    \sL_n^\dagger = - \bL_n, \hspace{1 cm} \bL^\dagger_n = - \sL_n
\end{equation}
which are \textit{not} the usual adjoint equations familiar from 2D CFT. We will now show that this is remedied by the $\rsw$ adjoint.

From the action of $\sR$ on the Lorentz generators\,,
\begin{equation}
  \begin{split}
    \sR \J_1 \sR &= - \J_1\,, \\
    \sR \K_1 \sR &= + \K_1\,,
  \end{split}
\quad \quad \quad
  \begin{split}
    \sR \J_2 \sR &= - \J_2\,, \\
    \sR \K_2 \sR &= + \K_2\,,
  \end{split}
\quad \quad \quad
  \begin{split}
    \sR \J_3 \sR &= + \J_3\,, \\
    \sR \K_3 \sR &= - \K_3\,,
  \end{split}
\end{equation}
we have the equations
\begin{align}\label{R_Ln}
\sR \sL_n \sR = - \bL_{-n}\,, \hspace{1 cm} \sR \bL_n \sR = - \sL_{-n}\,.
\end{align}
Moreover, from the equations~\eqref{primary_Ln_1},~\eqref{primary_Ln_2},~\eqref{shadow_Ln_1},~\eqref{shadow_Ln_2}, it is simple to verify that
$(\mathsf{SW}) \sL_{n} (\mathsf{SW}) = \sL_n$ and $(\mathsf{SW}) \bL_{n} (\mathsf{SW}) = \bL_n$.\footnote{The combined operator $\mathsf{SW}$ can be expressed as an integral transform in spacetime. See Appendix \ref{app_sc}.} This implies that
\begin{equation}\label{rsw_adj_Ln}
\begin{split}
    \sL_n^\drsw = \sL_{-n}\,, \hspace{1 cm} \bL_n^\drsw = \bL_{-n}\,,
\end{split}
\end{equation}
which are the usual adjoint equations familiar from 2D CFT. Note the above formula provides a proof for equation \eqref{rsw_descendants}.

The adjoint equations \eqref{rsw_adj_Ln} are highly useful: they completely determine $\rsw$ inner products of all descendants in $\mathcal{C}$ and $\widehat{\mathcal{C}}$. In addition, the momentum operators $\sP_\mu$ can be used to create a recursion relation between inner products of states with fixed conformal weights, e.g.~between $( \psi^+_{\Delta, N}, \psi^+_{\Delta, N})_{\rsw}$ and $( \psi^+_{\Delta+1, N}, \psi^+_{\Delta+1, N})_{\rsw}$. This computation is done in Appendix \ref{app_matrix_elts}.

The resulting inner products are given by
\begin{align}\label{matrix_elts_eq}
    \left( \sL_{-1}^{n_1} \bL_{-1}^{\bar{n}_1} \psi^\pm_{\Delta_1,N}, \sL_{-1}^{n_2} \bL_{-1}^{\bar{n}_2} \psi^\pm_{\Delta_2,N}
    \right)_{\rsw} &= \pm \delta_{n_1,n_2} \delta_{\bar{n}_1, \bar{n}_2} \delta_{\Delta_1, \Delta_2} M_{-\Delta_1, n_1, \bar{n}_1} \\
    \left( \sL_{-1}^{n_1} \bL_{-1}^{\bar{n}_1} \psi^\pm_{\Delta_1,N}, \sL_{-1}^{n_2} \bL_{-1}^{\bar{n}_2} \psi^\mp_{\Delta_2,N}
    \right)_{\rsw} &= 0
\end{align}
where
\begin{equation}\label{matrix_elts_const}
\begin{split}
    M_{\ell, n, \bar{n}} = -(2 \pi)^2 (-1)^{n + \bar{n}} \frac{n!\,  \ell!}{(\ell - n)!} \frac{\bar{n}! \, \ell!}{(\ell - \bar{n})!}\,.
\end{split}
\end{equation}
These $M_{\ell, n, \bar{n}}$ coefficients will appear in the following section where we provide our decomposition of the Wightman function.

\section{Decomposition of the Wightman function}
\label{sec_decomp}

Now we show how to decompose the Wightman function $G_+(X,Y)$ into elements of our basis $\mathcal{B}$ defined in~\eqref{E:Cset1},~\eqref{E:Cset2},~\eqref{basis_1}.  Writing the Wightman function as
\begin{equation}
G_+(X, Y) = \frac{1}{4\pi^2} \frac{1}{(X_+ - Y_-)^2}\,,
\end{equation}
we have
\begin{equation}\label{main_result}
\boxed{\frac{1}{4\pi^2} \frac{1}{(X_+ - Y_-)^2} = \sum_{\Delta \in \mathbb{Z}_{\leq 0}} \sum_{n,\bar{n} = 0}^{-\Delta}  \frac{\sL^n_{-1} \bL^{\bar{n}}_{-1} \psi_{\Delta,N}(X_+) \; \sL^n_{1} \, \bL^{\bar{n}}_{1} \,\widetilde{\psi_{2-\Delta,S}}(Y_-) }{M_{-\Delta,n,\bar{n}}}}
\end{equation}
This formula is the  main result of this paper. It establishes that $\mathcal{B}$ is complete in the sense we can use the basis to construct a kernel which evolves initial data of the massless Klein-Gordon equation to arbitrary times anywhere in the bulk.  The formula~\eqref{main_result} is a flat-space analog of the quasinormal mode de Sitter analysis in~\cite{Jafferis:2013qia}.

To prove equation~\eqref{main_result}, we first observe that the right-hand side is Poincar\'{e} invariant (see Appendix~\ref{app_poincare} for a proof).  As such, it is sufficient to check that \eqref{main_result} holds for two points of the form $X^\mu = (X^0,0,0,X^3)$ and $Y^\mu = (Y^0,0,0,Y^3)$.  Temporarily deferring our application of the $i\epsilon$ prescription for clarity, we can use the identities
\begin{align}\label{descendant_primary_plane}
\sL^n_{-1} \bL^{\bar{n}}_{-1} \psi_{\Delta,N}(X^0,0,0,X^3) &= \begin{dcases} \frac{n! \, (-\Delta)!}{(-\Delta - n)!}  (X^0 - X^3)^{-\Delta-n}(X^0 + X^3)^n &\text{for } n = \bar{n} \\
0 &\text{for } n \neq \bar{n} \end{dcases} \\ \nonumber \\
\sL^n_{1}\, \bL^{\bar{n}}_{1} \,\widetilde{\psi_{2-\Delta,S}}(Y^0, 0,0,Y^3) &= \begin{dcases} \frac{n! \, (-\Delta)!}{(-\Delta - n)!} (Y^0 + Y^3)^{-n-1}(Y^0 - Y^3)^{n+\Delta-1} &\text{for } n = \bar{n} \\ 0 &\text{for } n \neq \bar{n}
\end{dcases}\,,
\end{align}
which are proven in Appendix \ref{app_plane}, to find
\begin{align}
&\sum_{\Delta \in \mathbb{Z}_{\leq 0}} \sum_{n,\bar{n} = 0}^{-\Delta}  \frac{\sL^n_{-1} \bL^{\bar{n}}_{-1} \psi_{\Delta,N}(X^0, 0,0,X^3) \; \;  \sL^n_{1} \, \bL^{\bar{n}}_{1} \,\widetilde{\psi_{2-\Delta,S}}(Y^0,0,0,Y^3) }{M_{-\Delta,n,\bar{n}}}  \nonumber \\ \label{sum}
& \qquad = -\frac{1}{4 \pi^2} \sum_{\Delta \in \mathbb{Z}_{\leq 0}} \sum_{n = 0}^{-\Delta}  (Y^0 - Y^3)^{-1} (Y^0 + Y^3)^{-1} \left( \frac{X^0 - X^3}{Y^0 - Y^3} \right)^{-\Delta - n}  \left( \frac{X^0 + X^3}{Y^0 + Y^3} \right)^{n} \\
& \qquad = \frac{1}{4\pi^2} \frac{1}{-(X^0-Y^0)^2 + (X^3-Y^3)^2}\,.
\end{align}
Upon implementing the $i \epsilon$ prescription by sending $X^0 \mapsto X^0 - i \epsilon$ and $Y^0 \mapsto Y^0 + i \epsilon$, our argument is completed. Strictly speaking, the sum in \eqref{sum} does not converge for all values of $X^0, X^3, Y^0, Y^3$, but can be extended to all values via analytic continuation.

\section{Discussion}
\label{sec_discussion}

We have provided a basis expansion for solutions to the massless Klein-Gordon equation where each element of the basis has integer conformal dimension with respect to $\mathfrak{sl}(2,\mathbb{C})$.  Along the way, we have discovered the utility of the $\rsw$ inner product, which is best viewed as a modification to the Klein-Gordon inner product. This inner product is highly reminiscent of the BPZ inner product from 2D CFT where the conformal dimensions are non-negative integers. Although in this paper we have focused on massless scalars, it should be possible to treat massive fields in a similar fashion, although the structure of the solution space is rather different (see e.g.~\cite{pasterski2017conformal}).  We anticipate that our analysis should most readily generalize to massless spin-1 and spin-2 fields, which may clarify certain features of soft photon and soft graviton theorems and resonate with~\cite{Freidel:2022skz}.

Perhaps most intriguing are the implications of our work to the program of celestial holography.  Quantizing the massless scalar field with respect to our mode expansions and accompanying $\rsw$ inner product should provide a useful template for the structure of celestial CFTs.  We note that inner products similar to the $\rsw$ one have already appeared in the celestial holography literature~\cite{Crawley:2021ivb}, and our work may help to elucidate the meaning of the inner product.

\subsection*{Acknowledgements}
We thank Adam Ball, Erin Crawley, Elizabeth Himwich, Y.T. Albert Law, Sruthi Narayanan, Atul Sharma, Adam Tropper, Hongbao Zhang, and especially Tianli Wang for useful conversations. JC is supported by a Junior Fellowship from the Harvard Society of Fellows. This work was supported in part by NSF grant PHY-2207659.

\appendix

\section{Descendants as representations of Lorentz and Poincar\'{e} groups}
\label{App:finitedim}

\subsection{Descendants at fixed $\Delta$ as a finite dimensional irrep of the Lorentz group}

At fixed conformal dimensional $\Delta = - \ell$ for $\ell \in \{0, 1, 2, \ldots\}$, there are $(\ell + 1)^2$ descendants of the form $\sL^n_{-1} \bL^{\bar{n}}_{-1} \psi^+_{-\ell, N}$ with $n$ and $\bar{n}$ ranging from $0$ to $\ell$. (If $n > \ell$ of $\bar{n} > \ell$, the descendant is 0.) In fact, these descendants can be understood as being built from two spin-$\ell/2$ representations of $\mathfrak{so}(3)$, which is also known as the finite dimensional $(\tfrac{\ell}{2}, \tfrac{\ell}{2})$ representation of the Lorentz group. The purpose of this Appendix is to explicitly give the correspondence and illustrate how the $\rsw$ inner product relates to the standard inner product on $\mathfrak{so}(3)$ representations.

The $\mathfrak{so}(3)$ generators $\mj_i$ for $i = 1,2,3$ have the commutation relations $[\mj_i, \mj_j] = i \, \epsilon_{ijk} \mj_k.$ Defining the raising and lowering operators $\mj_\pm = \mj_1 \pm i \, \mj_2$, we have
\begin{equation}
    [\mj_3, \mj_\pm] = \pm \mj_\pm, \hspace{0.75 cm}  [\mj_+, \mj_-] = 2 \mj_3.
\end{equation}
In the spin $j$ representation of $\mathfrak{so}(3)$, the generators act on the state vectors $\ket{m,j}$ via
\begin{align}
\mj_3 \ket{m, j} &= m \ket{m, j} \\
\mj_\pm \ket{m, j} &= \sqrt{(j \mp m)(j \pm m + 1)} \ket{m\pm 1, j} \label{jpm}
\end{align}
where $m = j, \ldots, -j$. Here we use the standard inner product $\braket{m_1,j_1}{m_2,j_2} = \delta_{m_1,m_2} \delta_{j_1,j_2}$. From \eqref{jpm}, the norm of the states $\mj_-^n \ket{j, j}$ and $\mj_+^n \ket{-j, j}$ can be computed to be
\begin{equation}\label{so3_norms}
    \bra{j,j} \mj_+^n \mj_-^n \ket{j,j} = \frac{n! (2j)!}{(2j - n)!}, \hspace{1 cm} \bra{-j,j} \mj_-^n \mj_+^n \ket{-j,j} = \frac{n! (2j)!}{(2j - n)!}.
\end{equation}

If we define a second, independent set of $\mathfrak{so}(3)$ generators denoted $\tj_{\pm}$ and $\tj_3$, we can reproduce the $\mathfrak{sl}(2, \mathbb{C})$ commutation relations \eqref{L_commutators} via the identifications
\begin{equation}\label{so3_sl2r_id}
\begin{split}
    \mj_+ &\leftrightarrow - \sL_{1}\,, \\
    \tj_+ &\leftrightarrow  \bL_{-1}\,,
\end{split}
\quad \quad
\begin{split}
    \mj_- &\leftrightarrow \sL_{-1}\,, \\
    \tj_- &\leftrightarrow - \bL_{+1}\,, \\
\end{split}
\quad \quad
\begin{split}
    \mj_3 &\leftrightarrow - \sL_0\,, \\
    \tj_3 &\leftrightarrow \bL_0\,.
\end{split}
\end{equation}
In fact, using \eqref{sl2c_generators}, the above identifications \eqref{so3_sl2r_id} are are equivalent to 
\begin{equation}
    \begin{split}
        \mj_i &\; \leftrightarrow \; \frac{1}{2} ( \K_i + i \J_i) \, , \\
        \tj_i & \;\leftrightarrow \; \frac{1}{2} ( -\K_i + i \J_i) .
    \end{split}
\end{equation}
This means that if $\mj_i$ and $\tj_i$ generate the $j$ and $\widetilde{j}$ representations of $\mathfrak{so}(3)$, then the states can also be understood as transforming in the $(j, \widetilde{j})$ representation of the Lorentz group, which has dimension $(2j + 1) \times (2 \widetilde{j} + 1)$. Clearly, to make contact with the $(\ell + 1) \times (\ell + 1)$ dimensional space of $\mathfrak{sl}(2, \mathbb{C})$ descendants, we must set
\begin{equation}
    j = \tilde{j} = \frac{\ell}{2}.
\end{equation}
Under this correspondence, the north and south pole primaries are the states
\begin{equation}
\begin{split}
    \psi^+_{-\ell, N} \; &\leftrightarrow \; \ket{\tfrac{\ell}{2}, \tfrac{\ell}{2}} \otimes \ket{-\tfrac{\ell}{2}, \tfrac{\ell}{2}} \, , \\
    \psi^+_{-\ell, S} \; &\leftrightarrow \; \ket{-\tfrac{\ell}{2}, \tfrac{\ell}{2}} \otimes \ket{\tfrac{\ell}{2}, \tfrac{\ell}{2}}\,. \\
\end{split}
\end{equation}
If we define the ``descendants'' of our tensored spin-$\ell/2$ Hilbert spaces via
\begin{equation}
    \ket{n; \bar{n}} \vcentcolon= (\mj_-)^n (\tj_+)^{\bar{n}} \ket{\tfrac{\ell}{2}, \tfrac{\ell}{2}} \otimes \ket{-\tfrac{\ell}{2},\tfrac{\ell}{2}} \,,
\end{equation}
then these correspond directly to our $\mathfrak{sl}(2,\mathbb{C})$ descendants via 
\begin{equation}
    \sL_{-1}^n \bL_{-1}^{\bar{n}} \psi^+_{-\ell, N} \; \leftrightarrow \; \ket{n; \bar{n}}.
\end{equation}

In fact, from equations \eqref{matrix_elts_eq}, \eqref{matrix_elts_const}, and \eqref{so3_norms}, we can see that the $\rsw$ inner product is equal to the standard inner product of our $\mathfrak{so}(3)$ representations with the operator $(-1)^{\mj_3+ \mj_3}$ inserted in between:
\begin{equation}
\begin{split}
    \big(\sL_{-1}^{n_1} \bL_{-1}^{\bar{n}_1} \psi^+_{-\ell, N}, \sL_{-1}^{n_2} \bL_{-1}^{\bar{n}_2} \psi^+_{-\ell, N} \big)_{\mathsf{RSW}} = (- 4 \pi^2) \bra{n_1, \bar{n}_1} (-1)^{\mj_3 + \tj_3} \ket{n_2, \bar{n}_2}.
\end{split}
\end{equation}
Because the tensored $\mathfrak{so}(3)$ representation decomposes as
\begin{equation}
    \tfrac{\ell}{2} \otimes \tfrac{\ell}{2} = \ell \oplus (\ell - 1) \oplus \ldots \oplus 1 \oplus 0\,,
\end{equation}
our states are in correspondence with the set of spherical harmonics with $j \leq \ell$, namely
\begin{equation}
    \left\{ Y_{jm}(\theta, \phi) \right\}_{\substack{\!\!\! j = 0,\ldots,\ell \\ m = j,\ldots,- j}}.
\end{equation}
Interestingly, this is the exact same construction one encounters in the BFSS matrix model for matrices of size $(\ell+1)$ via matrix regularization of functions on the sphere \cite{deWit:1988wri, Taylor:2001vb}.

\subsection{Discrete irreducible representations of the Poincar\'{e} group}

The role of the momentum operators in the Poincar\'{e} algebra is to interlace the $(\tfrac{\ell}{2}, \tfrac{\ell}{2})$ representations of the Lorentz group for different $\ell$. Interestingly, the Poincar\'{e} representations formed from the shadowed and unshadowed primaries are quite different. While one could replace $\psi^+_{-\ell, N} \to \widetilde{\psi^-_{2 + \ell, S}}$ in the previous Subsection and the essential representation theory of the Lorentz group would not change, this is not true when translations are brought into the mix.

Using the south pole pointing null vector $S^\mu = (1,0,0,-1)$, it is straightforward to compute
\begin{align}
    (S^\mu \sP_\mu) \psi^+_{-\ell, N} &= - 2 \ell \, \psi^+_{-(\ell - 1), N} \\
    (S^\mu \sP_\mu) \widetilde{\psi^-_{2 + \ell, S}} &=  2(\ell+1) \widetilde{\psi^-_{2+(\ell+1), S}}\,. 
\end{align}
Therefore, while the momentum operator sends $\ell \to \ell -1$ for the primary basis, it sends $\ell \to \ell + 1$ for the shadow basis. Schematically, this can be represented as
\begin{equation}
\begin{split}
    &\mathcal{C} : \hspace{1 cm} \emptyset \overset{\sP_\mu}{\longleftarrow} (0,0) \overset{\sP_\mu}{\longleftarrow} (\tfrac{1}{2}, \tfrac{1}{2}) \overset{\sP_\mu}{\longleftarrow} (1,1) \overset{\sP_\mu}{\longleftarrow} (\tfrac{3}{2}, \tfrac{3}{2}) \overset{\sP_\mu}{\longleftarrow} (2,2) \overset{\sP_\mu}{\longleftarrow} \cdots \\
    &\widehat{\mathcal{C}} : \hspace{2.04 cm}(0,0) \overset{\sP_\mu}{\longrightarrow} (\tfrac{1}{2}, \tfrac{1}{2}) \overset{\sP_\mu}{\longrightarrow} (1,1) \overset{\sP_\mu}{\longrightarrow} (\tfrac{3}{2}, \tfrac{3}{2}) \overset{\sP_\mu}{\longrightarrow} (2,2) \overset{\sP_\mu}{\longrightarrow} \cdots
\end{split}
\end{equation}
At a conceptual level, we can imagine $\mathcal{C}$ as being generated by acting the Poincar\'{e} algebra on a wavefunction with $\ell = \infty$ (or perhaps more precisely, we can generate more of $\mathcal{C}$ by acting the Poincar\'{e} algebra on wave functions with progressively large $\ell$), whereas $\widehat{\mathcal{C}}$ can be generated by instead using the wavefunction with $\ell = 0$.

\section{Wightman function review}\label{sec_wightman}

For pedagogical purposes, in this Appendix we review fundamental properties of the Wightman function.  We begin with a purely classical approach, and then comment on its manifestation in quantum field theory with modified inner products.

\subsection{Classical perspective}

Suppose we are given initial conditions $\phi(X^0,\vec{X})$, $\partial_0\phi(X^0,\vec{X})$ to the massless Klein-Gordon equation on a time slice with fixed $X^0$ and would like to evolve to some other time slice $Y^0$.  Then the kernel
\begin{equation}
K(X, Y) \vcentcolon= \frac{i}{4\pi}\frac{1}{|\vec{X}-\vec{Y}|} \Big(\delta(|\vec{X}-\vec{Y}| - (X^0-Y^0)) - \delta(|\vec{X}-\vec{Y}| + (X^0-Y^0))\Big)
\end{equation}
allows us to evolve our initial conditions by
\begin{equation}
\phi(Y) = \big(K(X, Y)\,, \phi(X)\big)_{\mathrm{KG},\,X}
\end{equation}
where the $X$ subscript on the Klein-Gordon inner product indicates an integration over $\vec{X}$ on a surface of constant $X^0$.  As such, if we can reproduce the kernel $K(X, Y)$ with our desired mode expansion, then we can fully solve the massless Klein-Gordon equation as a classical PDE.

The kernel $K(X,Y)$ can also be split up into its positive and negative frequency parts, as the so-called Wightman Green's functions $G^+(X,Y)$ and $G^-(X,Y)$,
\begin{equation}\label{Gpm_function}
    G^\pm(X,Y) = \frac{1}{(2 \pi)^2} \frac{1}{-(X^0 - Y^0 \mp i \epsilon)^2 + (\vec{X} - \vec{Y})^2 } = \frac{1}{(2 \pi)^2} \frac{1}{(X - Y)^2 \pm i \epsilon \, \mathrm{sgn}(X^0 - Y^0)}.
\end{equation}
It can be readily checked that
\begin{equation}
    K(X,Y) = G^+(X,Y) - G^-(X,Y).
\end{equation}

\subsection{Quantum perspective}

While $K(X,Y)$ and $G^\pm(X,Y)$ can be understood as purely classical objects, it is also worthwhile to understand how they are instantiated within quantum field theory.

Recall that the quantum creation and annihilation operators are defined via
\begin{align} \label{annihilation_def}
    \hat{a}(f)& \vcentcolon=  ( f, \hat \phi )_{\mathrm{KG}} \\ \label{creation_def}
    \hat{a}^\dagger(f)& = - ( f^*, \hat \phi )_{\mathrm{KG}} \\
    [\hat{a}(f), \hat{a}^\dagger(g)] &= (f,g)_{\mathrm{KG}}\,.
\end{align}
If $f$ is a positive frequency wave function, meaning it is a linear combination of plane waves like $e^{i k \cdot X}$ for $k^0 > 0$, then $\hat{a}(f)\ket{0} = 0$ and we say $\hat{a}^\dagger(f) \ket{0}$ is a single particle state with wave function $f$. If one has two such positive frequency wave functions, then the Klein-Gordon inner product coincides with the standard QFT inner product because $\bra{0} a(f_1) \hat{a}^\dagger(f_2) \ket{0} = (f_1, f_2)_{\rm KG}$. Furthermore, it is straightforward to check that the Klein-Gordon inner product of a postive frequency wave function with a negative frequency wave function is always $0$.

In some contexts (such as this paper), it is advantageous to make use of other norms besides the Klein-Gordon norm. Say one uses an invertible linear operator $\mathcal{O}$ to define the inner product
\begin{equation}
    \left(f, g\right)_{\mathcal{O}} \vcentcolon= (f, \mathcal{O} g)_{\rm KG}
\end{equation}
and suppose a basis of positive frequency wave functions $f_i$ diagonalizes this inner product via
\begin{equation}
\begin{split}
    (f_i, f_j)_{\mathcal{O}} &= \delta_{ij} M_i, \\
    (f_i^*, f_j^*)_{\mathcal{O}} &= -\delta_{ij} (M_i)^*, \\
    (f_i, f_j^*)_{\mathcal{O}} &= 0,
\end{split}
\end{equation}
for some constants $M_i$. (Of course, in this particular paper, $\mathcal{O} = \rsw$ and the $f_i$ are given by the $\sL_{-1}^n \bL_{-1}^{\bar{n}} \psi^+_{\Delta,N} \in \mathcal{C}$, but let us keep our discussion more general here.)

If $f_i$ is a \textit{complete} basis of positive frequency wave functions, and $(\mathcal{O} f_i)^*$ is a \textit{complete} basis of negative frequency wave functions, then using \eqref{annihilation_def} and \eqref{creation_def}  we can decompose the field operator $\hat{\phi}(X)$ as
\begin{equation} \label{mode_exp}
    \hat{\phi}(X) = \sum_i \frac{1}{M_i^*} \left(  f_i(X) \hat{a}(\mathcal{O} f_i) +  (\mathcal{O} f_i)^*(X) \hat{a}^\dagger(f_i) \right).
\end{equation}

Returning to the Green's functions, one can show via explicit computation that they are given by the non-time-ordered two point functions
\begin{equation}
\begin{split}
    K(X,Y) &= \bra{0} [\; \hat{\phi}(X), \, \hat{\phi}(Y) \; ] \ket{0}, \\
    G^+(X,Y) &= \bra{0} \hat{\phi}(X)  \hat{\phi}(Y)  \ket{0}, \\
    G^-(X,Y) &= \bra{0} \hat{\phi}(Y)   \hat{\phi}(X)  \ket{0}.
\end{split}
\end{equation}
In particular, we can now expand the two point function $\bra{0}\hat{\phi}(X) \hat{\phi}(Y)\ket{0}$ using the mode expansion \eqref{mode_exp}. The result is
\begin{equation}\label{G_decomp}
    G^+(X,Y) = \sum_i \frac{f_i(X) (\mathcal{O} f_i)^*(Y)}{M_i^*}\,.
\end{equation}
(Note that \eqref{main_result} in the main text is exactly this equation.) Equation \eqref{G_decomp} is useful for the following reason: if one is not sure whether or not the bases $f_i$ and $(\mathcal{O} f_i)^*$ are \textit{complete} bases of positive and negative frequency wave functions respectively, one can simply evaluate the right-hand side of the above equation and see if it matches the left-hand side given in \eqref{Gpm_function}. If they match, the proposed bases are indeed complete.

It turns out that the Wightman function $G^+(X,Y)$ can also be thought of as a decomposition of the identity operator for positive frequency wave functions, analogous to the expression $I = \sum_i \frac{\ket{i}\bra{i}}{\braket{i}}$. To see this, one must simply take the Klein-Gordon inner product of $G^+(X,Y)$ and the initial data by integrating $X$ over a time slice and treating $Y$ as a constant. This is easiest to see by acting it on the function $\mathcal{O} f_j(X)$ and getting $\mathcal{O} f_j(Y)$ back out:
\begin{equation}
\begin{split}
    \left( G^+(X,Y), \mathcal{O} f_j(X) \right)_{\mathrm{KG}, X} &= \sum_i \frac{\mathcal{O} f_i(Y)}{M_i} \left( f_i(X), \mathcal{O} f_j(X) \right)_{\mathrm{KG}, X} \\
    &= \mathcal{O} f_j(Y).
\end{split}
\end{equation}
Viewing a full positive frequency solution in spacetime as a single state in our Hilbert space, we see that $G^+(X,Y)$ is precisely the identity operator on positive frequency wave functions when it is convolved in the Klein-Gordon inner product.

\section{Proving Poincar\'{e} invariance of Wightman function decomposition}\label{app_poincare}
In this Appendix we will prove that the sum on the right-hand side of \eqref{main_result} is Poincar\'{e} invariant. The proof is essentially symbol pushing. We say a function $F(X,Y)$ is Poincar\'{e} invariant if
\begin{equation}\label{poincare_inv_finite}
    F(\Lambda X + v, \Lambda Y + v) = F(X, Y)
\end{equation}
for all Lorentz transformations $\Lambda$ and four-vectors $v$. If we define an arbitrary element $\mathsf{A}$ of the Poincar\'{e} algebra in terms of basis elements \eqref{poincare_generators} as
\begin{equation}
 \mathsf{A} = \theta^i \J_i + \phi^i \K_i + v^\mu \sP_\mu\,,
\end{equation}
then the infinitesimal version of \eqref{poincare_inv_finite} is
\begin{equation}
    \mathsf{A}_{X} F(X, Y) + \mathsf{A}_{Y} F(X, Y) = 0\,,
\end{equation}
where the subscripts in $\mathsf{A}_{X}$ and $\mathsf{A}_{Y}$ denote the differential operator $\mathsf{A}$ acting on the variable $X$ and $Y$ respectively.

Say we have a function $F$ which is defined by
\begin{equation}
F(X, Y) \vcentcolon= \sum_{i} \frac{f_i(X) g_i^*(Y)}{M_i^*}\,,
\end{equation}
where $g_i = ( \rsw ) f_i$ and $( f_i, g_j)_{\rm KG} = \delta_{ij}M_i$. We further define the matrices $a_i^{\; i'}$ and $b^{j'}_{\; j}$ via the equations
\begin{equation}
    \mathsf{A} f_i = \sum_k a_i^{\; k} f_{k}, \hspace{1 cm} \mathsf{A} g_j = \sum_k g_{k} b^{k}_{\; j}.
\end{equation}
Here we require that $\mathsf{A} f_i$ and $\mathsf{A} g_j$ can both be written as linear combinations of $f_i$'s and $g_j$'s respectively, (i.e., that the sets $f_i$ and $g_j$ are closed under the Poincar\'{e} algebra) but crucially we do not need to assume that the combined set of $f_i$'s and $g_j$'s are complete.

The Poincar\'{e} invariance of the Klein-Gordon inner product can be written as
\begin{equation}
( \mathsf{A} f_i, g_j)_{\rm KG} + (f_i, \mathsf{A} g_j)_{\rm KG} = 0
\end{equation}
where we have used $\mathsf{A}^\dagger = - \mathsf{A}$.  The above equation is equivalent to
\begin{equation}
    (a_i^{\; j})^* M_{j} + M_{i} b^{i}_{\; j} = 0.
\end{equation}
Complex conjugating the above equation and dividing by $M_i^* M_j^*$ gives
\begin{equation}
    \frac{1}{M_i^*} a_i^{\; j} + \frac{1}{M_j^*} (b^{i}_{\; j})^* = 0.
\end{equation}
Finally, we compute
\begin{align}
    \mathsf{A}_{X} F(X,Y) + \mathsf{A}_{Y} F(X,Y) &= \sum_i \frac{1}{M_i^*} \left( (\mathsf{A} f_i) (X) g^*_i(X) + f_i(X) (\mathsf{A} g_i)^*(Y) \right) \\
    &= \sum_{i,j} \frac{1}{M_i^*} ( a_i^{\; j} f_{j}(X) g^*_i(Y) + f_i(X) g^*_{j}(Y) (b^{j}_{\; i})^* ) \\
    &= \sum_{i,j} \left( \frac{1}{M_i^*} a_i^{\; j} + \frac{1}{M_j^*} (b^i_{\; j})^*\right) f_j(X) g_i^*(Y) \\
    &= 0
\end{align}
which concludes the proof that $F(X,Y)$ is Poincar\'{e} invariant.

\section{Computing $M_{\ell,n,\bar{n}}$ from Poincar\'{e} symmetry}\label{app_matrix_elts}

We will now show how the $M_{\ell,n,\bar{n}}$ coefficients of \eqref{matrix_elts_const} can be determined from Poincar\'{e} symmetry up to an overall multiplicative constant.

Before doing so, it is worth explaining why the the $\rsw$ inner product between $\sL^n_{-1} \bL^{\bar{n}}_{-1} \psi^\pm_{-\ell,N}$ and $\sL^{n'}_{-1} \bL^{\bar{n}'}_{-1} \psi^\pm_{-\ell',N}$ is only non-zero if $n = n'$, $\bar{n} = \bar{n}'$, and $\ell = \ell'$, as indicated in \eqref{matrix_elts_eq}. First, note that $\sL_{\pm 1}$ and $\bL_{\pm 1}$ are raising and lowering operators with respect to $\sL_0$ and $\bL_0$. This implies that $\sL^n_{-1} \bL^{\bar{n}}_{-1} \psi^\pm_{-\ell,N}$ is an eigenstate of $\sL_0$ and $\bL_0$ with eigenvalues $-\ell/2 + n$ and $-\ell/2 + \bar{n}$, respectively. Because $\sL_0^{\drsw} = \sL_0$  and $\bL_0^{\drsw} = \bL_0$, we have that $\sL_0$ and $\bL_0$ are self-adjoint with respect to the $\rsw$ inner product. By basic linear algebra, this means that wave functions which are eigenstates of $\sL_0$ and $\bL_0$ with different eigenvalues must have an inner product of 0. 

Let us now compute the recursion relation between $M_{\ell,n,\bar{n}}$ and $M_{\ell,n-1,\bar{n}}$.  We have
\begin{align}\label{Mn_recur}
    M_{\ell,n,\bar{n}} &= \left( \sL_{-1}^n \bL_{-1}^{\bar{n}} \psi^+_{-\ell,N}, \sL_{-1}^n \bL_{-1}^{\bar{n}} \psi^+_{-\ell,N} \right)_{\rsw} \\
    &= \left( \sL_{-1}^{n-1} \bL_{-1}^{\bar{n}} \psi^+_{-\ell,N}, \sL_1 \sL_{-1}^n \bL_{-1}^{\bar{n}} \psi^+_{-\ell,N} \right)_{\rsw} \nonumber \\
    &= 2 \big( (-\ell/2 + n - 1) + (-\ell/2 + n - 2) + \ldots + (-\ell/2)  \big) \left( \sL_{-1}^{n-1} \bL_{-1}^{\bar{n}} \psi^+_{-\ell,N}, \sL_{-1}^{n-1} \bL_{-1}^{\bar{n}} \psi^+_{-\ell,N} \right)_{\rsw} \nonumber \\
    &= n (n - \ell - 1) M_{\ell,n-1,\bar{n}}\,, \nonumber
\end{align}
where in the second line we used $\sL_{-1}^{\drsw} = \sL_1$, in the third line we commuted $\sL_1$ past each instance of $\sL_{-1}$ using $[\sL_1, \sL_{-1}] = 2 \sL_0$ until $\sL_1$ annihilated $\bL_{-1}^{\bar{n}} \psi^+_{-\ell,N}$ at the end, and along the way used that $\sL_{-1}^n \bL_{-1}^{\bar{n}} \psi^+_{-\ell,N}$ is an eigenstate of $\sL_0$ with eigenvalue $-\ell/2 + n$.

Likewise, we also have the recursion relation
\begin{equation}\label{Mnbar_recur}
M_{\ell,n,\bar{n}} = \bar{n}( \bar{n} - \ell - 1) M_{\ell, n, \bar{n} - 1}.
\end{equation}
Solving the recursion relations \eqref{Mn_recur} and \eqref{Mnbar_recur}, we have
\begin{equation}\label{recursion_1}
    M_{\ell,n,\bar{n}} = (-1)^{n + \bar{n}} \frac{n! \, \ell!}{(\ell - n)!} \frac{\bar{n}! \, \ell!}{(\ell - \bar{n})! }  M_{\ell,0,0}.
\end{equation}

Having made full use of Lorentz symmetry, we must use the momentum operators $\sP^\mu$ to create a recursion relation between $M_{\ell,0,0}$ and $M_{\ell-1,0,0}$. (Actually, one can compute $M_{\ell,0,0}$ for all $\ell$ directly from~\eqref{rsw_primary} (just set $z_1 = z_2 = 0$) finding $M_{\ell,0,0} = - (2 \pi)^2$, but it is nice give a symmetry argument as well.)

We need to use a translation operator which changes $\ell$. For simplicity, we use the specific translation operator $S^\mu \sP_\mu$, where $S^\mu = (1,0,0,-1)$ is the south pointing null vector 

On the primary states, we can directly compute
\begin{equation}\label{e5}
    (S^\mu \sP_\mu) \psi^\pm_{-\ell,N} = -2 \ell \, \psi^{\pm}_{-\ell + 1, N}.
\end{equation}
On the shadowed states $(\rsw) \psi^\pm_{-\ell+1,N} = \widetilde{\psi^\pm_{\ell+1,S}}$, we can also directly compute
\begin{equation}\label{e6}
    (S^\mu \sP_\mu) (\mathsf{RSW}) \psi^\pm_{-\ell+1,N} = 2 \ell \, (\mathsf{RSW}) \psi^\pm_{-\ell , N}.
\end{equation}
Using $(S^\mu \sP_\mu)^\dagger = - S^\mu \sP_\mu$, we find
\begin{align}
    ( (S^\mu \sP_\mu) \psi^+_{-\ell,N}, (\mathsf{RSW}) \psi^+_{-\ell+1,N} )_{\mathrm{KG}} &= - ( \psi^+_{-\ell,N},  (S^\mu \sP_\mu) (\mathsf{RSW}) \psi^+_{-\ell+1,N} )_{\mathrm{KG}}
\end{align}
which upon using \eqref{e5} and \eqref{e6} becomes\footnote{More precisely, the recursion relation is $\ell M_{\ell+1,0,0} = \ell M_{\ell,0,0}$. In order to divide by $\ell$ in the $\ell = 0$ case, we must take the limit $\ell \to 0$ and not plug in $\ell = 0$ directly.} 
\begin{equation}\label{recursion_2}
    M_{(\ell+1),0,0} = M_{\ell,0,0}\,.
\end{equation}
This is our new recursion relation. Combining \eqref{recursion_1} and \eqref{recursion_2}, we have
\begin{equation}
    M_{\ell,n,\bar{n}} =(-1)^{n + \bar{n}} \frac{n! \, \ell!}{(\ell - n)!} \frac{\bar{n}! \, \ell!}{(\ell - \bar{n})! } M_{0,0,0}
\end{equation}
which determines $M_{\ell,n,\bar{n}}$ up to a multiplicative constant which turns out to be $M_{0,0,0} = - (2 \pi)^2$.

\section{Expression for descendants on the $X^1 = X^2 = 0$ plane}\label{app_plane}

We set $\Delta = - \ell$ where $\ell \in \{0, 1, 2, \ldots\}$. We begin by considering which wave functions $\sL^n_{-1} \bL^{\bar{n}}_{-1} \psi_{-\ell,N}(X)$ are nonzero when $X^1 = X^2 = 0$. Noting that $\psi_{-\ell,N}(X) = (X^0 - X^3)^\ell$, all we must do is study the repeated action of $\sL_{-1}$ and $\bL_{-1}$ on $(X^0 - X^3)$. It is straightforward to compute
\begin{equation}
\begin{split}
    \sL_{-1} (X^0 - X^3) &= - X^1 + i X^2, \\
    \bL_{-1} (X^0 - X^3) &= - X^1 - i X^2,
\end{split}
\quad \quad \quad
\begin{split}
    (\sL_{-1})^2 (X^0 - X^3) &= 0, \\
    (\bL_{-1})^2 (X^0 - X^3) &= 0,
\end{split}
\end{equation}
\begin{equation}
    \sL_{-1} \bL_{-1} (X^0 - X^3) = X^0 + X^3.
\end{equation}
The above equations imply that, if one is to compute $\sL^n_{-1} \bL^{\bar{n}}_{-1} (X^0 - X^3)^\ell$ using the product rule of differentiation, the only terms which will be non-zero when $X^1 = X^2 = 0$ will be terms where each factor of $(X^0 - X^3)$ which was acted on by one $\bL_{-1}$ was then subsequently acted on by one $\sL_{-1}$. (Note that this can only hold when $n = \bar{n}$.) By a combinatorial argument, the number of such terms will be $n! \, \ell! / (\ell - n)!$, so we have
\begin{equation}
(\sL^n_{-1} \bL^{\bar{n}}_{-1} \psi_{-\ell,N})(X^0, 0,0,X^3) = \begin{dcases} \frac{n! \, \ell!}{(\ell - n)!}  (X^0 - X^3)^{\ell-n}(X^0 + X^3)^n & n = \bar{n} \\
0 & n \neq \bar{n} \end{dcases}.
\end{equation}

Let us now do the analogous computation for the shadowed modes. We can once again compute
\begin{equation}
\begin{split}
    \sL_{1} (X^0 + X^3) &= X^1 + i X^2, \\
    \bL_{1} (X^0 + X^3) &= X^1 - i X^2,
\end{split}
\quad \quad \quad
\begin{split}
    (\sL_{1})^2 (X^0 + X^3) &= 0, \\
    (\bL_{1})^2 (X^0 + X^3) &= 0,
\end{split}
\end{equation}
\begin{equation}
    \sL_{1} \bL_{1} (X^0 + X^3) = X^0 - X^3.
\end{equation}
Using the equation $\widetilde{\psi_{2+\ell,S}} = (-X^\mu X_\mu)^{-\ell-1} (X^0 + X^3)^\ell$ and noting that $\sL_{1} (X^\mu X_\mu) = \bL_{1} (X^\mu X_\mu) = 0$, we can repeat our previous analysis to obtain

\begin{equation}
(\sL^n_{1} \, \bL^{\bar{n}}_{1} \,\widetilde{\psi_{2+\ell,S}} )(X^0,0,0,X^3) = \begin{dcases} \frac{n! \, \ell!}{(\ell - n)!} (X^0 + X^3)^{-n-1}(X^0 - X^3)^{n-\ell-1} & n = \bar{n} \\ 0 & n \neq \bar{n}
\end{dcases}.
\end{equation}

We conclude with the interesting observation that if we take $n = \bar{n} = \ell$, our north pole primaries are sent to the south pole and vice versa:
\begin{equation}
    \sL^{\ell}_{-1} \bL^{\ell}_{-1} \psi_{-\ell,N} = (\ell!)^2 \psi_{-\ell,S} \hspace{1 cm} \sL^{\ell}_{1} \, \bL^{\ell}_{1} \,\widetilde{\psi_{2+\ell,S}} = (\ell!)^2 \widetilde{\psi_{2+\ell,N}}.
\end{equation}

\section{Inner product integrals} \label{app_inner_prod}

\subsection{Klein-Gordon inner product of primaries}\label{app_kg}

In this Appendix we will derive the formula for the Klein-Gordon inner product of two conformal primary modes
\begin{align} \label{to_prove_1}
    \left( \psi^\pm_{-\nn, \qone}, \psi^\pm_{-\nm, \qtwo} \right)_{\mathrm{KG}} &= \mp \frac{ (2 \pi)^3 }{|\nn - \nm|} \delta^2(z_1 - z_2) \delta_{\nn+\nm, -2} \\
    \left( \psi^\pm_{-\nn, \qone}, \psi^\mp_{-\nm, \qtwo} \right)_{\mathrm{KG}} &= 0\label{to_prove_2}
\end{align}
for $\nn \in \{0, 1, 2, \ldots\}$, $\nm \in \{-2, -3, -4, \ldots \}$ or $\nm \in \{0, 1, 2, \ldots\}$, $\nn \in \{-2, -3, -4, \ldots \}$. Strictly speaking, these inner products involve integrals that must be computed using a regularization which shall be given, and whichever of $\nn$ or $\nm$ is in the set $\{ 0, 1, 2, \ldots \}$ must be evaluated in the limit that it approaches said integer.

\subsubsection{$\pm \pm$ case}

We start with the $++$ case. The Klein-Gordon inner product can be written as
\begin{align}\label{kg_I_2}
    \left( \psi_{-\nn, \qone}^+, \psi_{-\nm, \qtwo}^+ \right)_{\mathrm{KG}} = i \, \nm \, q^0(\ztwo) \; \mathcal{I}^{++}(\nn, \nm-1) - i \, \nn \, q^0(\zone) \; \mathcal{I}^{++}(\nn-1, \nm)
\end{align}
where the integrals $\mathcal{I^{\pm \pm}}(\nn, \nm)$ are defined on the $X^0 = 0$ slice via
\begin{align}
    \mathcal{I}^{++}(\nn, \nm) &= \int d^3 X (- \vqone \cdot \vec{X} + i \epsilon)^{\nn} (- \vqtwo \cdot \vec{X} - i \epsilon)^{\nm} \\
    \mathcal{I}^{--}(\nn, \nm) &= \int d^3 X (- \vqone \cdot \vec{X} - i \epsilon)^{\nn} (- \vqtwo \cdot \vec{X} + i \epsilon)^{\nm} \\
    \mathcal{I}^{+-}(\nn, \nm) &= \int d^3 X (- \vqone \cdot \vec{X} + i \epsilon)^{\nn} (- \vqtwo \cdot \vec{X} + i \epsilon)^{\nm} \\
    \mathcal{I}^{-+}(\nn, \nm) &= \int d^3 X (- \vqone \cdot \vec{X} - i \epsilon)^{\nn} (- \vqtwo \cdot \vec{X} - i \epsilon)^{\nm}
\end{align}
where the four integrals have different combinations of signs of the $i \epsilon$'s.

In order to compute these integrals, we will employ the use of the Riemann-Liouville fractional derivative
\begin{equation}\label{riemann}
    {}_\lambda D^\np_\alpha f(\alpha) = \begin{dcases} 
    \left( \frac{d}{d\alpha}\right)^\nl \frac{1}{\Gamma(\nl - \np)} \int_\lambda^\alpha (\alpha - \alpha')^{\nl - \np - 1 } f(\alpha') d \alpha' & \hspace{1 cm} \text{ if } \np > 0, \np \notin \mathbb{Z} \\
    \frac{1}{\Gamma(|\np|)} \int_\lambda^\alpha (\alpha - \alpha')^{|\np| - 1 } f(\alpha') d \alpha' & \hspace{1 cm} \text{ if } \np < 0
    \end{dcases}
\end{equation}
where $\nl \in \mathbb{Z}_{\geq 0}$. This operator mimics both differentiation and integration depending on the sign of $\np$. The reader should be aware that ${}_\lambda D^{\np}_\alpha$ is ill-defined when $\np$ is a non-negative integer $\np = 0,1,2, \ldots$. If we ever wish to calculate ${}_\lambda D^{\np}_\alpha f(\alpha)$ when $\np$ is one of these integers, we must do so with a limit $\lim_{\np' \to \np}( {}_\lambda D^{\np'}_\alpha f(\alpha) )$. This observation will become important later on.

An important property the operator ${}_\lambda D^{\np}_\alpha$ satisfies is
\begin{equation}\label{RL_exp}
\begin{split}
    \lim_{\alpha \to 0^+}  \big[ {}_{+\infty} D^{\np}_\alpha ( e^{i A \alpha} ) \big] = (i A)^{\np} & \hspace{0.5 cm} \text{ if } \Im{A} > 0 \text{ for all } {\np} \in \mathbb{R} \setminus \{0,1,2, \ldots\} \\
    \lim_{\beta \to 0^-}  \big[ {}_{-\infty} D^{\np}_\beta ( e^{i A \beta} ) \big] = (i A)^{\np} & \hspace{0.5 cm} \text{ if } \Im{A} < 0 \text{ for all } {\np} \in \mathbb{R} \setminus \{0,1,2, \ldots\}.
\end{split}
\end{equation}
The prescribed direction of the limits $\alpha \to 0^+$ and $\beta \to 0^-$ in the above equations, which depend on the sign of $\Im{A}$, is needed in the forthcoming steps so that ${}_{+\infty} D^{\np}_\alpha$ and ${}_{-\infty} D^{\np}_\beta$ reproduce expected derivative properties when the real part of $A$ is integrated over.\footnote{Here we elaborate on the direction of the limit of $\alpha \to 0$. Without loss of generality, take $\Im{A} = \epsilon > 0$ where $\epsilon$ is small. Consider the integral $I(\alpha) = \int_{-\infty + i \epsilon}^{+\infty + i \epsilon} \frac{d A}{2 \pi} e^{i A \alpha} = \delta(\alpha)$. We want ${}_{+\infty} D^{\np}_\alpha I(\alpha)$ to mimic the expected behavior of $\partial_\alpha^{\np} \delta(\alpha)$ for $p > 0$, in particular that $\lim_{\alpha \to 0} \partial_\alpha^{\np} \delta(\alpha) = 0$. So, which direction for the limit of $\alpha \to 0$ should we take for ${}_{+\infty} D^{\np}_\alpha I(\alpha)$ to be guaranteed to be zero, for $\np \notin \mathbb{Z}$? Inspecting ${}_{+\infty} D^{\np}_\alpha I(\alpha) = \int_{-\infty + i \epsilon}^{+\infty + i \epsilon} \frac{d A}{2 \pi} ( i A)^{\np} e^{i A \alpha}$, we have a branch cut from the fractional exponent $A^{\np}$ for $A$ on the negative real axis. This branch cut is avoided by the contour of integration which runs from $- \infty + i \epsilon$ to $+ \infty + i \epsilon$. In order to close the contour of the $A$ integral in the upper half plane and show the integral is zero, we need for $\alpha > 0$. Therefore, the correct direction of the $\alpha \to 0$ limit is $\alpha \to 0^+$ for $\Im{A} > 0$. Likewise, we require $\alpha \to 0^-$ for $\Im{A} < 0$.
}

We now use the Riemann-Liouville fractional derivative and \eqref{RL_exp} to express $\mathcal{I}^{++}(\nn, \nm)$ via
\begin{equation}\label{Ipm_intermediate}
    \mathcal{I}^{++}(\nn,\nm) =  (-i)^{\nn + \nm} \lim_{\alpha \to 0^+}  ({}_{+\infty}D^{\nn }_{\alpha}) \lim_{\beta \to 0^-} ({}_{- \infty}D^{\nm}_{\beta}) \int d^3 X e^{i \alpha (- \vqone \cdot \vec{X} + i \epsilon)} e^{i \beta(- \vqtwo \cdot \vec{X} - i \epsilon)}
\end{equation}
where the integral on the right-hand side of the above equation evaluates to
\begin{align}\
    \int d^3 X e^{i \alpha (- \vqone \cdot \vec{X} + i \epsilon)} e^{i \beta(- \vqtwo \cdot \vec{X} - i \epsilon)} &= (2 \pi)^3  \delta^3(\alpha \vec{q}(\zone) + \beta \vec{q}(\ztwo) ) e^{\epsilon(\beta - \alpha)} \\
    &= - \frac{(2 \pi)^3}{4 \alpha \beta} \frac{\delta^2(z_1 - z_2)}{q^0(\zone)} \delta(\alpha + \beta)  e^{\epsilon(\beta - \alpha)}. \label{feyn_intermediate}
\end{align}
Plugging \eqref{feyn_intermediate} into \eqref{Ipm_intermediate} gives
\begin{equation}\label{Ipm_exp}
    \mathcal{I}^{++}(\nn,\nm) = - (-i)^{\nn + \nm} \frac{(2 \pi)^3}{4} \frac{\delta^2(z_1 - z_2)}{q^0(z_1, \bz_1)}  \lim_{\alpha \to 0^+}  ({}_{+\infty}D^{\nn }_{\alpha}) \lim_{\beta \to 0^-} ({}_{- \infty}D^{\nm}_{\beta}) \frac{e^{\epsilon(\beta - \alpha)}}{\alpha \beta} \delta(\alpha + \beta).
\end{equation}
One may then compute
\begin{align}
    \lim_{\alpha \to 0^+} ({}_{+\infty} D^{\nn}_{\alpha}) \; \lim_{\beta \to 0^-} ({}_{- \infty} D^{\nm}_\beta)  \; \frac{e^{\epsilon(\beta - \alpha)}}{\alpha \beta} \delta( \alpha + \beta) & = \lim_{\alpha \to 0^+} ({}_{+\infty} D^{\nn}_{\alpha}) \;  \frac{-1}{\Gamma(-\nm)} \alpha^{-\nm-3} e^{-2 \epsilon \alpha}  \Theta(\alpha) \\
    &=  \frac{\Gamma(-3-\nn-\nm)}{\Gamma(-\nn) \Gamma(-\nm)} (-1)^{\nn+1}  (2\epsilon)^{3 + \nn + \nm}. \label{b15}
\end{align}
Plugging \eqref{b15} into \eqref{Ipm_exp} gives
\begin{equation}\label{d16}
    \mathcal{I}^{++}(\nn,\nm) = i^{\nn-\nm} \frac{(2 \pi)^3}{4} \frac{\delta^2(z_1 - z_2)}{q^0(z_1, \bz_1)} \frac{\Gamma(-3-\nn-\nm)}{\Gamma(-\nn) \Gamma(-\nm)}  (2\epsilon)^{3 + \nn + \nm}
\end{equation}
and plugging \eqref{d16} this into \eqref{kg_I_2} we get
\begin{equation}\label{b17}
    \left( \psi_{-\nn, \qone}^+, \psi_{-\nm, \qtwo}^+ \right)_{\mathrm{KG}} = \frac{(2 \pi)^3}{2} \delta^2(z_1 - z_2)  i^{\nn-\nm} \frac{\Gamma(-2 - \nn - \nm)}{\Gamma(- \nn ) \Gamma(- \nm )} (2 \epsilon)^{2 + \nn + \nm}.
\end{equation}

In this final step, we must recall that the Riemann-Liouville fractional derivative is not defined at positive integer values, and a limit must be taken. Let us first assume that $\nn \in \{ 0, 1, 2, \ldots\}$ and $\nm \in \{ -2, -3, -4, \ldots\}$. While we can plug $\nm$ directly into the above formula, $\nn$ must be taken as a limit approaching an integer. If we then take the $\epsilon \to 0^+$ limit afterwards, we get
\begin{equation}
\begin{split}
    \lim_{\epsilon \to 0^+} \lim_{\nn' \to \nn} i^{\nn'-\nm} \frac{\Gamma(-2 - \nn' - \nm)}{\Gamma(- \nn' ) \Gamma(- \nm)} (2 \epsilon)^{2 + \nn'+\nm} = \frac{-1}{\nn + 1}\delta_{\nn+\nm, -2} 
    \hspace{0.5 cm} \text{ for } \hspace{0.5 cm} \substack{\!\!\!\!\!\!\!\!\!\!\!\!\!\!\! \nn \in \{ 0, 1, 2, \ldots\} \\ \nm \in \{ -2, -3, -4, \ldots\}}.
\end{split}
\end{equation}
Likewise, if we instead take $n \in \{ -2, -3, -4, \ldots\}$ and $m \in \{0, 1, 2, \ldots\}$, we get
\begin{equation}
    \lim_{\epsilon \to 0^+} \lim_{\nm' \to \nm} i^{\nn-\nm'} \frac{\Gamma(-2 - \nn - \nm')}{\Gamma(- \nn ) \Gamma(- \nm')} (2 \epsilon)^{2 + \nn + \nm'} = \frac{-1}{\nm + 1}\delta_{\nn + \nm, -2} \hspace{0.5 cm}\text{ for } \hspace{0.5 cm} \substack{\nn \in \{ -2, -3, -4, \ldots\} \\\!\!\!\!\!\!\!\!\!\!\!\!\!\!\!   \nm \in \{ 0,1,2, \ldots\}}.
\end{equation}
Plugging the above two equations into \eqref{b17}, we can write the combined result as
\begin{equation}
    \left( \psi_{-\nn, \qone}^+, \psi_{-\nm, \qtwo}^+ \right)_{\mathrm{KG}} = -\frac{(2 \pi)^3}{|\nn -\nm|} \delta^2(z_1 - z_2)  \delta_{\nn+\nm,-2}
\end{equation}
as desired, which holds for both the case $\nn \in \{0, 1, 2, \ldots\}$, $\nm \in \{ -2, -3, -4, \ldots\}$ as well as the case $\nn \in \{ -2, -3, -4, \ldots\}$, $\nm \in \{0, 1, 2, \ldots\}$. This completes the computation of the $++$ case of \eqref{to_prove_1}. 

Note that from the identity $(f^*, g^*)_{\rm KG} = - (f,g)_{\rm KG}^*$ we obtain the $-$$-$ case of \eqref{to_prove_1} as well.

\subsubsection{$\pm \mp$ case}

We start with the $+-$ case of \eqref{to_prove_2}. The Klein-Gordon inner product can be written as
\begin{align}\label{kg_I_pm_mp}
    \left( \psi_{-\nn, \qone}^+, \psi_{-\nm, \qtwo}^- \right)_{\mathrm{KG}} = i \, \nm \, q^0(\ztwo) \; \mathcal{I}^{+-}(\nn, \nm-1) - i \, \nn \, q^0(\zone) \; \mathcal{I}^{+-}(\nn-1, \nm)
\end{align}
and following a similar logic as the last section, the integral $\mathcal{I}^{+-}(\nn,\nm)$ is given by
\begin{equation}
    \mathcal{I}^{+-}(\nn,\nm) = - (-i)^{\nn + \nm} \frac{(2 \pi)^3}{4} \frac{\delta^2(z_1 - z_2)}{q^0(z_1, \bz_1)}  \lim_{\alpha \to 0^+}  ({}_{+\infty}D^{\nn }_{\alpha}) \lim_{\beta \to 0^+} ({}_{+ \infty}D^{\nm}_{\beta}) \frac{e^{\epsilon(-\alpha - \beta)}}{\alpha \beta} \delta(\alpha + \beta).
\end{equation}
One may then compute
\begin{align}
    \lim_{\alpha \to 0^+}  ({}_{+\infty}D^{\nn }_{\alpha}) \lim_{\beta \to 0^+} ({}_{+ \infty}D^{\nm}_{\beta}) \; \frac{e^{\epsilon(-\alpha - \beta)}}{\alpha \beta} \delta(\alpha + \beta) &= \lim_{\alpha \to 0^+}  ({}_{+\infty}D^{\nn }_{\alpha}) \; \frac{1}{\Gamma(-\nm)}  \alpha^{-\nm-3} \Theta(- \alpha)
    = 0\,.
\end{align}
which implies
\begin{equation}
    \mathcal{I}^{+-}(\nn,\nm) = 0.
\end{equation}
Accordingly, we have
\begin{align}\label{kg_I}
    \left( \psi_{-\nn, \qone}^+, \psi_{-\nm, \qtwo}^- \right)_{\mathrm{KG}} = 0\,.
\end{align}
which completes the proof of the $+-$ case of \eqref{to_prove_2}.
Once again, from the identity $(f^*, g^*)_{\rm KG} = - (f,g)_{\rm KG}^*$ we obtain the $-$$+$ case of \eqref{to_prove_2} as well.

\subsection{$\rsw$ inner product}\label{app_shadow}

In this Appendix we will compute the $\rsw$ inner products of the conformal primary wave functions using the previously calculated Klein-Gordon inner products \eqref{to_prove_1} and \eqref{to_prove_2}. For $\nn, \nm \in \{ 0,1,2, \ldots\}$, we will find that the $\rsw$ inner products are
\begin{align}\label{rsw_to_prove_1}
    \left( \psi^\pm_{-\nn, q(z_1, \bz_1)}, (\mathsf{RSW}) \psi^\pm_{-\nm, q(z_2, \bz_2)} \right)_{\mathrm{KG}} &= \mp ( 2 \pi)^2 |1 - z_1 \bz_2 |^{2 \nn} \delta_{\nn,\nm} \\
    \left( \psi^\pm_{-\nn, q(z_1, \bz_1)}, (\mathsf{RSW}) \psi^\mp_{-\nm, q(z_2, \bz_2)} \right)_{\mathrm{KG}} &= 0\,.
    \label{rsw_to_prove_2}
\end{align}
Equation \eqref{rsw_to_prove_2} follows directly from \eqref{to_prove_2}. We now carry out the computation of \eqref{rsw_to_prove_1} by acting $\mathsf{W}$, $\sS$, and $\sR$ successively on the second term in the Klein-Gordon inner product, using \eqref{to_prove_1} as our starting point. Applying $\mathsf{W}$ to \eqref{to_prove_1} just sends $m \to -2 - m$ and we get
\begin{equation}
    \left( \psi^\pm_{-\nn, q(z_1, \bz_1)}, (\mathsf{W}) \psi^\pm_{-\nm, q(z_2, \bz_2)} \right)_{\mathrm{KG}} = \mp \frac{(2 \pi)^3}{2(\nn + 1)} \delta^2(z_1 - z_2) \delta_{\nn,\nm}.
\end{equation}
We then insert the shadow operator $\sS$ into the above equation. Using the definition of $\sS$ given in \eqref{S_def}, it is straightforward to compute
\begin{equation}\label{int_sw}
    \left( \psi^\pm_{-\nn, q(z_1, \bar{z}_1)}, (\mathsf{SW}) \psi^\pm_{-\nm, q(z_2, \bar{z}_2)} \right)_{\mathrm{KG}} = \mp ( 2 \pi)^2 |z_1 - z_2|^{2 \nn} \delta_{\nn,\nm}.
\end{equation}
Now we must finally act with $\sR$. Here we note the identity
\begin{equation}
    \sR q(z, \bz) = |z|^2 \, q(1/\bz, 1/z)
\end{equation}
from which we compute
\begin{equation} \label{r_int}
    (\mathsf{R})  \widetilde{\psi}_{2+\nm, q(z,\bz)} = \frac{(- (\sR q(z, \bz)) \cdot X)^\nm}{(-X^\mu X_\mu)^{1 + \nm}} = |z|^{2 \nm} \widetilde{\psi^\pm}_{2+\nm, q(1/\bz,1/z)}.
\end{equation}
If we insert $\sR$ into \eqref{int_sw} and use \eqref{r_int}, we then get
\begin{equation}\label{rsw_primaries}
\begin{split}
    \left( \psi^\pm_{-\nn, q(z_1, \bz_1)}, (\mathsf{RSW}) \psi^\pm_{-\nm, q(z_2, \bz_2)} \right)_{\mathrm{KG}} &= \mp ( 2 \pi)^2 |z_2|^{2 \nm} |z_1 - 1/\bz_2|^{2 \nn} \delta_{\nn,\nm} \\
    &= \mp ( 2 \pi)^2 |1 - z_1 \bz_2 |^{2 \nn} \delta_{\nn,\nm}
\end{split}
\end{equation}
which completes the proof of \eqref{rsw_to_prove_1}.

\subsection{Relation to previous literature}

Previous literature \cite{pasterski2017conformal} has used another normalization convention for the conformal primary states $\psi^\pm_{\Delta, q}$ defined in \eqref{E:twokinds1}, differing by a factor of $(\mp i)^\Delta \Gamma(\Delta)$. The Klein-Gordon inner product of such states is proportional to the Dirac delta distribution $\delta( i (2 -\Delta_1^* - \Delta_2 ))$.

We stripped the $(\mp i)^\Delta \Gamma(\Delta)$ prefactor off of our definition because we are studying integer modes and $\Gamma(\Delta)$ diverges at $\Delta \in \{ 0, -1, -2, \ldots \}$. This is why we had to re-compute the Klein-Gordon inner products from scratch using different integration techniques. However, if we na\"{i}vely divide the Klein-Gordon inner product in \cite{pasterski2017conformal} by these divergent $\Gamma(\Delta)$ factors, we can reproduce our KG equations \eqref{to_prove_1} and \eqref{to_prove_2} if we use the ad-hoc replacement rules $\delta( i (2 + \nn + \nm )) \mapsto \delta(0) \delta_{\nn + \nm, -2}$ and $\frac{\delta(0)}{\Gamma(0)} \mapsto \frac{1}{2 \pi}$.

\section{$\mathsf{SW}$ as a spacetime integral transform}\label{app_sc}

The shadow operator $\mathsf{S}$ is defined as an integral over the momentum coordinate $q(z, \bar{z})$ in \eqref{S_def}. However, the combined operator $\mathsf{SW}$ is actually a well-known integral transform over the spacetime coordinate $X$ on the null cone. This integral transform is known among mathematicians and is used to construct the inner product for the discrete series representation of the Lorentz group, see for instance Vol.~2, Sections 9.2.7 and 9.2.8 of \cite{klimyk1995representations}.

We define $\widetilde{\mathsf{S}}$ via the position integral over a 2D slice of the null cone (denoted $\int d^2 Y$, see \cite{simmons2014projectors} for a review of these kinds of embedding space integrals) via
\begin{equation}
    \widetilde{\mathsf{S}} (- q \cdot X )^{-\Delta} \vcentcolon= \frac{ 1 - \Delta}{\pi}\int d^2 Y (- 2 X \cdot Y)^{\Delta-2} \frac{1}{(-q \cdot Y)^\Delta}
\end{equation}
Using the Feynman parameterization trick and the fundamental integral $\int d^2 Y ( - 2 Y \cdot Z )^{-2} = \pi / (- Z^\mu Z_\mu)$, one can compute
\begin{equation}
    \widetilde{\mathsf{S}} (- q \cdot X)^{-\Delta} = \frac{(-X^\mu X_\mu )^{\Delta-1}}{(-q \cdot X)^\Delta}
\end{equation}
which demonstrates that $\widetilde{\mathsf{S}} = \mathsf{SW}$. It should also be noted that $\widetilde{\mathsf{S}}$ is also equal to the so-called ``Kelvin transform'' $\mathcal{K}$, which acts on functions $f_\Delta(X)$ of scaling dimension $\Delta$ as
\begin{equation}
    (\mathcal{K} f_{\Delta})(X) = (-X^\mu X_\mu)^{(\Delta - 2)/2} f_{\Delta}\left(\frac{X}{  (-X^\mu X_\mu)^{1/2}}\right).
\end{equation}

\bibliography{refs}
\bibliographystyle{JHEP}

\end{document}